\documentclass[12pt, letterpaper]{article}
\usepackage[pass]{geometry}
\usepackage[utf8]{inputenc}
\usepackage{authblk}
\usepackage{amsmath}
\usepackage{amsfonts}
\usepackage{amsthm}
\usepackage{amssymb}
\usepackage{setspace}
\usepackage{csquotes}
\usepackage{fullpage}\usepackage{graphicx}
\usepackage{caption}
\usepackage{subcaption}
\usepackage[table,xcdraw]{xcolor}
\usepackage{bbm}
\newtheorem{definition}{Definition}

\usepackage[
    backend=biber,
    style=nature,
  ]{biblatex}

\usepackage{graphicx}
\usepackage{caption}
\usepackage{subcaption}
\usepackage{comment}

\theoremstyle{remark}

\title{Evaluating Binary Outcome Classifiers Estimated from Survey Data}

\author[1]{Adway S. Wadekar \thanks{adway.wadekar@duke.edu}}
\author[1]{Jerome P. Reiter}
\affil[1]{Department of Statistical Science, Duke University}

\date{July 2024}

\begin{document}

\maketitle

\begin{abstract}
Surveys are commonly used to facilitate research in epidemiology, health, and the social and behavioral sciences. Often, these surveys are not simple random samples, and respondents are given weights reflecting their probability of selection into the survey. 
We show that using survey weights 
can be beneficial for evaluating the quality of predictive models when splitting data into training and test sets.  In particular, we characterize model assessment statistics, such as sensitivity and specificity, as finite population quantities and compute survey-weighted estimates of these quantities with test data comprising a random subset of the original data. Using simulations with data from the National Survey on Drug Use and Health and the National Comorbidity Survey, we  show that unweighted metrics estimated with sample test data  can misrepresent population performance, but weighted metrics appropriately adjust for the complex sampling design. We also show that this conclusion holds for models trained using upsampling for mitigating class imbalance. The results suggest that weighted metrics should be used when evaluating performance on test data derived from complex surveys.
\end{abstract}


\section*{Introduction}\label{}
Empirical research in diverse fields, particularly within epidemiology and the social and behavioral sciences,
relies on data collected via sample surveys \cite{safdarResearchMethodsHealthcare2016, glock1967survey, wright2010survey, fabic2012systematic}. 
Frequently, and especially for large-scale surveys run by government agencies, the surveys use complex design features like stratification, clustering, and unequal probability sampling. 
Each individual has a survey weight that, loosely, indicates how many individuals it represents in the population  
\cite{pfeffermannUseSamplingWeights1996}. 
Generally, the weight for any individual equals the inverse of that individual's probability of selection into the survey, possibly with some multiplicative adjustment to account for survey nonresponse or to ensure weighted estimates of certain quantities are close to known population totals \cite{mansourniaInverseProbabilityWeighting2016}. 

It is well known that analysts should account for the survey design when estimating finite population quantities; for example, analysts can estimate population means or totals using the unbiased estimator of Horvitz and Thompson \cite{horvitzGeneralizationSamplingReplacement1952}. 
Analysts also can account for survey designs in explanatory models like linear and logistic regressions, for example, 
by using survey-weighted versions of those models  
\cite{dumouchelUsingSampleSurvey1983, reiterAnalyticalModelingComplex2005, faiellaUseSurveyWeights2010}. Often, however, analysts using data from complex survey designs do not use survey weights when fitting explanatory models  \cite{bellUseDesignEffects2012}. Indeed, there is debate as to whether or not survey weights are  needed in model-based analyses at all \cite{bollenAreSurveyWeights2016}. 

Independent of this debate, epidemiologic, public health, and other social science researchers increasingly are using predictive modeling as an alternative to explanatory analyses \cite{naimietal, jiangSupervisedMachineLearning2020a, bzdok, bennetetal}. As examples, classification techniques have been used to predict substance users in a sample of Mexican children \cite{vazquezInnovativeIdentificationSubstance2020a}; machine learning methods have been used to identify adults at risk for opioid or other substance use disorder \cite{wadekarPsychosocialApproachPredicting2020a, wadekarUnderstandingOpioidUse2020a}; 
and, predictive modeling has been used to identify veterans at risk for suicidal ideation after a year of service  \cite{borowskiFirstYearMilitary2022a}.
With the predictive analysis paradigm, the primary goal is to develop models 
that can be used to identify individuals in the population who might be at risk for a disease or other outcome, as opposed to identifying factors that might be associated with the outcome. To do so, researchers have expanded beyond parametric models to using classification and regression trees (CART) and random forests as well as black-box models like neural networks, sometimes  in ensemble \cite{hastie2009elements}. 


As with parametric models, the use of survey weights in the development of predictive models is idiosyncratic. Some analysts use them and some do not; see below 
for examples. 
The incorporation of survey weights can be complicated by some predictive modeling strategies.  For instance, in data with imbalanced distributions of outcomes, it is common practice to go against the grain of representative survey weighting and up-weight, down-weight, or even synthetically generate observations to balance the outcome distribution in the training data to assist with model development~\cite{jiangSupervisedMachineLearning2020a}.  Similarly, and of primary relevance here, some analysts use survey weights when assessing the performance of predictive models and others do not \cite{wieczorek, iparragirre2022estimation, iparragirre:stat}.

In this article, we investigate the role of survey weights in the testing stage of predictive analysis for binary outcome classifiers. 
We do so for a common data science pipeline: construct a model on a random subsample of the survey data using, when necessary, procedures like  downsampling or SMOTE \cite{chawlaSMOTESyntheticMinority2002a} to adjust for class imbalance, use the model to predict outcomes for a holdout sample from the survey, and compute estimates of sensitivity and specificity using the holdout sample.  We propose finite population quantities corresponding to the sensitivity and specificity that one would obtain by applying the predictive model to all individuals in the population, and we define survey-weighted estimators of those finite population quantities.  Using simulation studies, we show that these 
survey-weighted estimators 
can approximate the corresponding  finite population quantities more accurately than unweighted estimators. 
These findings apply for any predictive analysis on survey data, including those that use pre-processing steps like upsampling in model development.  R code for computing the survey-weighted metrics is available at \url{https://github.com/adway/survey-weights}.

The remainder of the article is organized as follows. First, we briefly review survey weights and  summarize related work from the machine learning literature.  
Second, we present the survey-weighted estimators for predictive model evaluation.  Third, we present results of  simulation studies used to evaluate these estimators.  These simulations use data from the National Survey on Drug Use and Health~\cite{u.s.departmentofhealthandhumanservicessubstanceabuseandmentalhealthservicesadministrationcenterforbehavioralhealthstatisticsandqualityNationalSurveyDrug2021} as well as data from the National Comorbidity Survey 2004 replication~\cite{kesslerNationalComorbiditySurvey2004}.  Finally, we conclude by summarizing the findings. An online supplement provides supporting material.  

\section*{Background}
\label{sec:background}

We first review key concepts from  survey sampling. We then review literature related to weighting in machine learning methods.

\subsection*{Survey Weights}
\label{sec: weight-lit}

Let $P$ be a finite population comprising $N$ individuals.  Each individual $i=1, \dots, N$ is measured {on $h$ survey variables}, which we write as $\mathbf{z}_i = (z_{i1}, \dots, z_{ih})$.  Thus, $P = \{\mathbf{z}_i: i = 1, \dots,  N\}$.  Let $D$ comprise $n$ randomly sampled individuals from $P$; these are the data used for predictive modeling. For $i=1, \dots, N$, we define the survey inclusion indicator $d_i=1$ when individual $i$ is sampled for the survey and $d_i=0$ when not. 
Each individual has some nonzero probability $\pi_i$ of being included in $D$, {that is, $Pr(d_i=1)=\pi_i > 0$.}  The values of $\pi_i$ are determined by the randomization scheme used to select the sample.  

{Many survey analyses make use of the inverse probability of selection, $w_i=1/\pi_i$, i.e.,  the  
survey weight. Each $w_i$ can be interpreted as the number of individuals in $P$ represented by individual $i$.}  
Some national surveys adjust survey weights to $w_i = c_i(1/\pi_i)$, where $c_i$ is some multiplicative factor, for example, to account for nonresponse, reduce standard errors of survey-weighted estimates, or to {calibrate to auxiliary totals using 
raking    \cite{little1986survey, deville1992calibration, brick2013unit,  silva2004properties, phipps2012analyzing, kott2016calibration}. }



For the case of $w_i=1/\pi_i$ for all $i$,  Horvitz and Thompson present an  estimator of population totals that can be applied for any sampling design~\cite{horvitzGeneralizationSamplingReplacement1952}. For any variable $z_k$, let $t_{z_k} = \sum_{i =1}^N z_{ik}$ be the total of $z_k$ in $P$.  We seek to estimate $t_{z_k}$ from $D$. The estimator is 
\begin{equation}
\hat{t}_{z_k} = \sum_{i \in D} w_iz_{ik}.\label{htestimator}
\end{equation}
Horvitz and Thompson  prove that $\mathbb{E}[\hat{t}_{z_k}] = t_{z_k}$, that is, the estimator in \eqref{htestimator} is unbiased for any survey design. {In these expectations, all individuals' $w_i$ and $z_{ik}$ are considered fixed features of that individual.} The random variables are the indicators $d_i$ of who is randomly selected to be in the sample.  The estimator in \eqref{htestimator} also can be unbiased when the $w_i$ are adjusted weights under certain conditions, for example, on the reasons why individuals do not respond to the survey \cite{littlerubin}.  We assume that the analyst uses whatever survey weights $w_i$ are available in the data files.


\subsection*{Weights in Predictive Modeling}
\label{sec: lit}

{An array of literature discusses weighting and non-representative data in machine learning frameworks for model training.  This literature generally does not address survey weights for computing evaluation metrics like sensitivity and specificity. }
Rather, much of it focuses on using analyst-specified weights to train models accurately, with an ultimate goal of mitigating the class imbalance problem \cite{huangLearningDeepRepresentation2016, byrdWhatEffectImportance2019, xuUnderstandingRoleImportance2021}. For example, one study investigates how models should be altered when training on data from one source and testing on another~\cite{steingrimsson2023}, and another discusses the use of weights to train and evaluate models when outcome labels are assigned not with certainty but with a degree of confidence~\cite{keilwagen}.  

Zadrozny shows that machine learning classifiers fit to nonrepresentative data can result in biased estimates of classification~\cite{zadroznyLearningEvaluatingClassifiers2004}.  Additionally,  Zadrozny  presents a theoretical approach to adjust machine learning model fits for nonrepresentative data. The basic idea is to introduce weights to the loss function used for model fitting. Each individual's  weight is the ratio of an  estimated marginal probability that the individual is in the training data over an estimated conditional probability that the individual is in the training data, given some analyst-specified set of variables.  
Although similar in objectives, {our contributions differ from Zadrozny in that we use weights that account for the probability of selection into the survey and subsequently the subsample of cases in the evaluation set.}  {These weights differ from ratios of estimated marginal and conditional probabilities that do not necessarily account for the two-stage sampling process.   
We also present metrics specifically to estimate sensitivity and specificity in complex samples rather than arbitrary loss functions with analyst-specified weights.} Finally, we perform repeated sampling simulations using training and testing data with genuine complex survey data, comparing both weighted and unweighted performance metrics against defined, population-level results.


Turning to the roles of complex survey weights in predictive modeling, Yao et al. present survey-weighted estimators for area-under-the-curve (AUC) metrics~\cite{yao:li}.  Their approach is targeted at estimating the AUC of a fitted model estimated with all the training data; they do not discuss splitting into training and evaluation data. They also do not consider procedures like downsampling to mitigate class imbalance.  Tsujimoto
also studies survey-weighted estimates of AUCs of this form~\cite{tsujimoto}.

MacNell et al. examine the use of weights when estimating and evaluating gradient boosting classifiers~\cite{macnellImplementingMachineLearning2023}. They treat a survey-weighted version of the classifier coupled with survey-weighted estimates of performance as a gold standard, and compare the performance of unweighted classifiers against this standard.
Wieczorek et al. examine the use of weights when evaluating classifiers using $k$-fold 
cross validation  
and ultimately recommend using survey-weighted estimates for evaluations~\cite{wieczorek}.  
While the topics and findings of these two papers are similar to ours, our work contributes to the literature in ways not addressed by these papers. Principally, we formally define finite population versions of sensitivity and specificity for general classifiers, whereas these two papers leave such quantities unspecified.   
Formal definitions of finite population quantities enable theoretical arguments for why estimators of them are (or are not) consistent, regardless of the classifier used to make predictions. In addition, these quantities facilitate considerations of variances of the estimators and for computation of estimators of the variances. 
They also facilitate benchmarking the weighted and unweighted performance metrics against a truth, as opposed to comparing against one another. Thus, we can  
assess which, if any, of the metrics is a reliable estimator of population-level accuracy. 

\section*{Methodology}
\label{sec:methods}

We begin by defining some standard performance metrics that are commonly used in assessing the predictive performance of classifiers. Let $TP$ refer to the number of true positives predicted by a model, and let $TN$ refer to the number of true negatives predicted by a model. Similarly, let $FP$ and $FN$ be the number of false positives and false negatives predicted by the model, respectively. 
\begin{definition}
\label{defn:uwsens}
	Sensitivity is the percentage of observations with a positive outcome that the predictive model identifies correctly. 
	\begin{align}
		Sensitivity = \frac{TP}{TP + FN}.
	\end{align}
\end{definition}
\begin{definition}
\label{defn:uwspec}
Specificity is the percentage of observations with a negative outcome that the predictive model identifies correctly. 
	\begin{align}
		Specificity = \frac{TN}{TN + FP}.
	\end{align}	
\end{definition}

To compute sensitivity and specificity, often analysts  partition the collected data into a training set $D_t$ and test set $D_e$, where $D=(D_t, D_e)$.  We assume the analyst determines $D_e$ (and thus $D_t$) by taking a simple random sample of $D$, although the analyst may choose some other design \cite{wieczorek}.  For $i=1, \dots, N$, let $e_i= 1$ when individual $i$ is selected into $D_e$ and $e_i=0$ otherwise. Note that $e_i = 1$ only if $d_i = 1$, and $e_i=0$ automatically for individuals not in $D$.  


We conceive of values of sensitivity and specificity computed from $D_e$ as estimates of quantities defined over $P$.  Our goal, then, is to obtain approximately unbiased estimates of these population quantities, as we now describe. To begin, we slightly modify the notation for the survey variables.  For $i=1, \dots, N$,  we write $\mathbf{z}_i = (\mathbf{x}_i, y_i)$, where $y_i$ represents an outcome of interest and $\mathbf{x}_i$ represents a set of predictor variables of interest.  Here, we assume that each $y_i \in \{0, 1\}$.  The analyst fits a predictive model using $D_t$ that can take any $\mathbf{x}_i$ and produce a predicted value $\hat{y}_i$.  We refer to this model as $M$ so that 
$\hat{y}_i = M(\mathbf{x}_i)$.

Suppose that we could apply $M$ to every individual in $P$. Then, we would have a collection of predictions 
$\hat{y}_i$ for all $N$ individuals in $P$. As a result, for $i=1, \dots, N$, we could define each individual in $P$ as a true positive, true negative, false positive, or false negative under $M$ using the indicator variables,
\begin{align}
	TP_i &= \mathbbm{1}(\hat{y}_i = y_i \textrm{ and } y_i = 1) \label{tpi}\\
	TN_i &= \mathbbm{1}(\hat{y}_i = y_i \textrm{ and } y_i=  0) \label{tni}\\
	FP_i &= \mathbbm{1}(\hat{y}_i \neq y_i \textrm{ and } y_i = 0) \label{fpi}\\
	FN_i &= \mathbbm{1}(\hat{y}_i \neq y_i \textrm{ and } y_i = 1). \label{fni} 
\end{align}   
Here, $\mathbbm{1}(\cdot)=1$ when the event inside the parentheses is true, and $\mathbbm{1}(\cdot)=0$ otherwise. For a fixed $M$, each individual in $P$ can have only one of these indicators equal to 1. 

Using these indicators, we define the finite population quantities,
\begin{align}
    N_{TP} = \sum_{i=1}^N TP_i \label{TP}\\
    N_{TN} = \sum_{i=1}^N TN_i \label{TN}\\
    N_{FP} = \sum_{i=1}^N FP_i \label{FP}\\
    N_{FP} = \sum_{i=1}^N FN_i. \label{FN}
\end{align}
Thus, we define a finite population sensitivity under $M$ as $N_{TP}/(N_{TP}+N_{FN})$ and a finite population specificity as $N_{TN}/(N_{TN}+N_{FP})$.  
 
We next consider how to estimate these finite population quantities.  In doing so, we conceive of each $\hat{y}_i$ as a fixed characteristic of individual $i$, where $i=1, \dots, N$.  Put another way,  we pretend that the analyst would like to apply $M$ and hence compute  $\hat{y}_i$ for all individuals in $P$, and that the analyst would not use some other $M$ with different samples of $D$ or $D_t$. Obviously, this may not be true in practice.  If the analyst had obtained a different $M$ from another $D$ or $D_t$, they would want to evaluate the quality of those values of $M(\mathbf{x}_i)$  for the individuals in $P$. However, conceiving of $M$ as fixed simplifies understanding of the finite population quantities in \eqref{TP}--\eqref{FN}.

Because we use random sampling to create $D_e$, we can use  Horvitz-Thompson estimators for \eqref{TP}--\eqref{FN}. We need the probability $\pi_i^*$ that any individual $i$ is in $D_e$, i.e.,  $\pi_i^* = P(e_i=1)$.  For $i=1, \dots, N$, we have
\label{def: compound-prob}
	\begin {align} 
	\pi_i^*  =   \mathbb{P}(d_i = 1) \mathbb{P}(e_i = 1 \mid d_i =1) = \pi_i \pi_{ei}.
	\end {align}	
Typically, we obtain $D_e$ as a simple random sample of $n_e$ individuals from the $n$ in $D$, so that $\mathbb{P}(e_i = 1 \mid d_i =1) = \pi_{ei}=n_e/n$. Thus, for each individual $i=1, \dots, n$ in $D$, we define 
\label{def: compound-weight}
	\begin {align} 
	w_i^* =  w_i(n/n_e),
	\end {align}	
where $w_i$ is the  original survey weight. 


With these weights, we can use  Horvitz-Thompson estimators for \eqref{TP}--\eqref{FN}.  We have
\begin{align}
	\widehat{N}_{TP} = \sum_{i \in D_e} w_i^* TP_i \label{TPest}\\
 \widehat{N}_{TN} = \sum_{i \in D_e} w_i^* TN_i \label{TNest}\\
 \widehat{N}_{FP} = \sum_{i \in D_e} w_i^* FP_i\label{FPest}\\
 \widehat{N}_{FN} = \sum_{i \in D_e} w_i^* FN_i.\label{FNest}
\end{align}

The expectations of the estimators in \eqref{TPest}--\eqref{FNest} are equal to the corresponding quantities in \eqref{TP}--\eqref{FN}, assuming that $M$ is fixed.  We show that this is the case for $\widehat{N}_{TP}$; proofs for the other estimators are similar. Assuming for simplicity that each $w_i=1/\pi_i$, we have 
\begin{align}
	\mathbb{E}\left[\widehat{N}_{TP}\right] &= \mathbb{E}\left[ \mathbb{E}\left[\widehat{N}_{TP}|D\right]\right]  = \mathbb{E}\left[ \mathbb{E}\left[\sum_{i \in D_e} {w_i^*} TP_i | D\right]\right] =  \mathbb{E}\left[\mathbb{E}\left[\sum_{i \in D_e} {(1/(\pi_i\pi_{ei}))} TP_i|D\right]\right]\\ 
 &=  \mathbb{E}\left[\sum_{i \in D} \mathbb{E}\left[(1/(\pi_i\pi_{ei})) e_i TP_i|D\right]\right] =  \mathbb{E}\left[\sum_{i \in D} (1/\pi_i) TP_i\right] \\
	&=  \mathbb{E}\left[\sum_{i \in P} (1/\pi_i) d_i TP_i\right] = \sum_{i \in P} TP_i
\end{align}
since  $\mathbb{E}(e_i|D) = \pi_{ei}$ and $\mathbb{E}(d_i) = \pi_i$ for all $i=1, \dots, N$.
The  random variables in these expectations are the indicators $d_i$ and $e_i$; all $TP_i$ are considered fixed attributes of the individual $i$, for fixed $M$.


We plug in the estimates from \eqref{TPest}--\eqref{FNest}  to obtain the estimates of the sensitivity (SN) and specificity (SP).  We have  
	\begin{align}
		\widehat{SN} &= \frac{\widehat{N}_{TP}}{\widehat{N}_{TP} + \widehat{N}_{FN}} \label{SNhat} \\
		\widehat{SP} &= \frac{\widehat{N}_{TN}}{\widehat{N}_{TN} + \widehat{N}_{FP}}. \label{SPhat} 
	\end{align}
Ratios of Horvitz-Thompson estimators are consistent for the corresponding population ratios  
\cite{lohrSamplingDesignAnalysis2021}.  Estimates of standard errors of ratio estimators can be obtained using Taylor series approximations.  These are available in survey sampling software like the ``survey'' package in R \cite{lumley}.  We note that survey-weighted estimates of sensitivity and specificity like \eqref{SNhat} and \eqref{SPhat} {have been proposed previously \cite{iparragirre2022estimation, iparragirre2023estimation}.}


\section*{Simulation Studies}
\label{sec:sims}
We present three simulation studies to examine the repeated sampling properties of $\widehat{SN}$ and $\widehat{SP}$.  
In the first, we construct a population using data from the National Survey on Drug Use and Health (NSDUH), from which we take stratified samples.
In the second and third, we use data from the NSDUH and the 2004 replication of the National Comorbidity Survey (NCS-R). The data derive from existing public use files that we downloaded from government agency websites, and as such, no institutional review board review was needed. We use the adjusted survey weights available on the data files. We do not know the values of $\mathbf{z}_i$ for records not in the surveys, so that we only can note differences in the values of  weighted and unweighted estimates. 

\subsection*{Simulation using constructed population from NSDUH}
\label{sub:complex}

We treat data from the NSDUH as a population, and take samples from it.  Here, we are motivated by previous analyses, which use data from the NSDUH to predict opioid use disorder~\cite{wadekarPredictingOpioidUse2019}.  Here, logistic regression, random forests, and decision trees are used as classifiers, and unweighted estimates of performance metrics are reported. 

Specifically, we combine data from the 2017, 2018, and 2019 editions of the NSDUH  to construct a population of $N=116,761$ individuals. {We ensure that all $N$ individuals have values for all variables of interest.  This necessitates discarding 11,558 individuals with partially complete data from the original data files.} 
As the outcome variable, 
we set $y_i=1$ if individual $i$ has any illicit drug use in the past year, and $y_i=0$ otherwise.  Marginally, 31,807 (27.2\%) people have illicit drug use and 84,954 (72.8\%) people do not.
As the predictors, we use sex, race, age category, income, {an indicator of any mental illness within the last year}, education level, self-reported health rating, employment status, age of first use of alcohol, age of first use of marijuana, parole status, probation status, approached by someone using drugs, obesity, whether heroin is fairly or very easy to obtain, perception of great risk trying heroin once or twice, and perception of great risk trying heroin weekly.

\begin{figure}
\centering
  \includegraphics[scale=0.5]{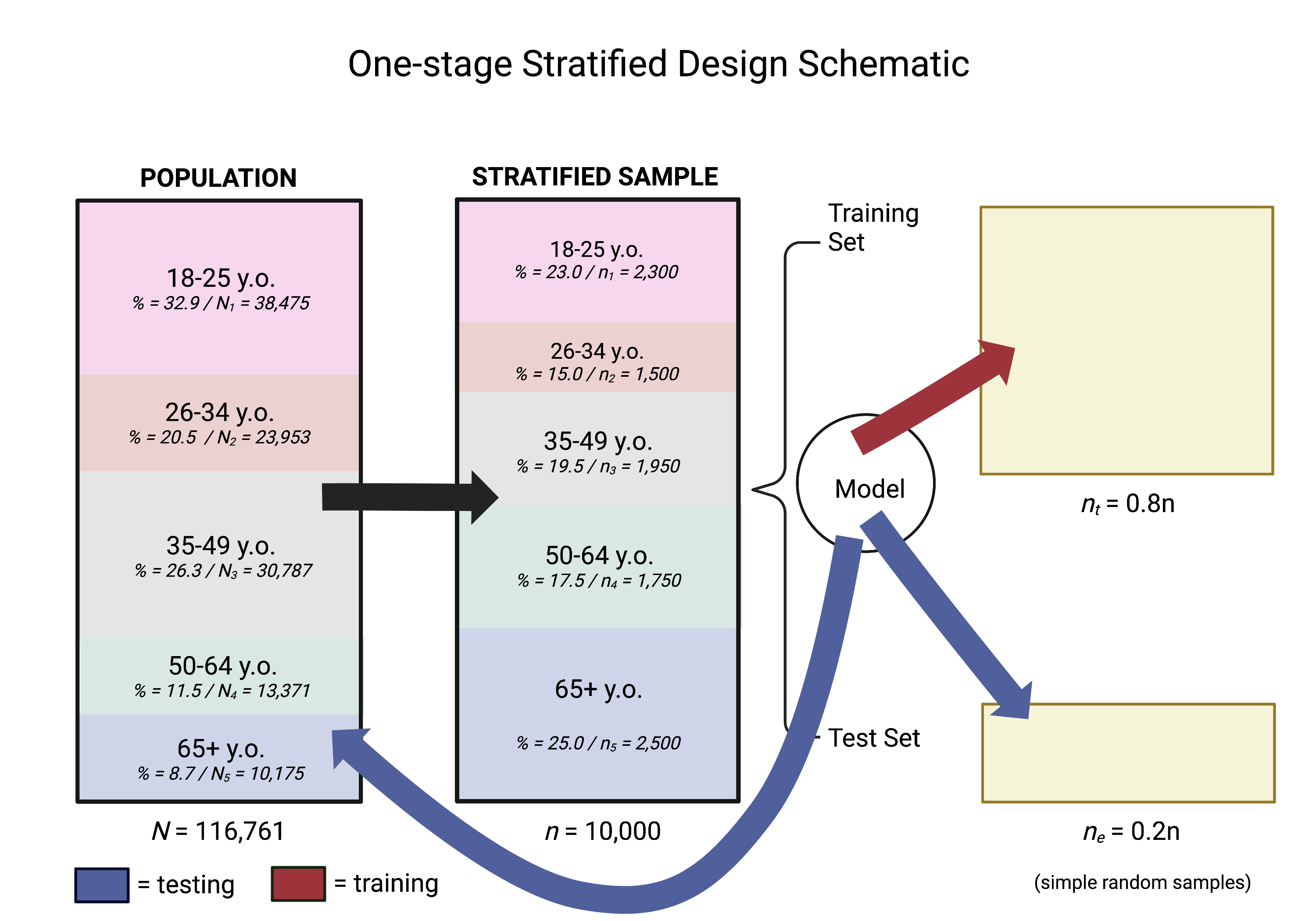}
  \caption{Sampling design schematic for the stratified sampling simulation.}
  \label{fig:sds}
\end{figure}

Figure~\ref{fig:sds} displays a  schematic for 
any one run of the simulation.  First, we sample $n=10,000$ records from the constructed $P$ according to a stratified design. The strata are age categories defined in the NSDUH data files, namely  18--25 years, 26--34 years, 35--49 years, 50--64 years, and over 65 years. Labeling the strata in increasing age order, the strata sizes in $P$ are $N_1 = 38,475, N_2 = 23953, N_3 = 30,787, N_4 = 13,371,$ and $N_5 = 10,175.$  As sample sizes for $D$, we use $n_1 = 2,300, n_2 = 1,500, n_3 = 1,950, n_4 = 1,750$ and $n_5 = 2,500$.  Thus, the sampling rates are not proportional to the population sizes. {We oversample the oldest and youngest groups to demonstrate a design where ignoring weights could be problematic for evaluation.}

We split the stratified sample $D$ into training and test sets.  We train three classifiers on an 80\% sample of $D$: a baseline logistic regression classifier, a random forest classifier that does not adjust for class imbalance in the outcome variable, and a random forest classifier that does adjust for class imbalance using a standard upsampling approach \cite{japkowicz2002class}. Specifically, we randomly sample and duplicate observations in the minority class in the 80\% training set until the number of observations associated with each outcome is balanced.  The resulting data are used to fit the classifier.  The test set, $D_e$, remains unbalanced. For the random forests, we use the ``randomForest'' package in R \cite{liawRandomForest}. We use the default splitting Gini impurity splitting criteria and set the number of trees in the forest to 1,000. 


We repeat the process of sampling $D$ and fitting the classifiers 500 times. In each run, we 
apply the classifiers trained on $D_t$ to all $N$ records in $P$, including those in $D$. In this way, in each run we can compute $(N_{TP}, N_{TN}, N_{FP}, F_{FN})$, since we know $y_i$ for all units in $P$.  These population quantities allow us to compute $SN$ and $SP$ for each fitted classifier.  Although the $10,000$ values in $D$ are included in each population quantity in each run, the population quantities are determined mainly by the estimated classifier's performance on the $106,761$ out-of-sample cases. By averaging the population quantities over the 500 runs, we also approximate their sampling distributions over repeated draws of $D$.

In each of the 500 runs, we also compute $\widehat{SN}$ using \eqref{SNhat} and $\widehat{SP}$ using \eqref{SPhat} for the $D_e$ in that run.  For comparisons, we  compute unweighted estimates of sensitivity and specificity using the usual expressions applied to $D_e$ as in Definitions \ref{defn:uwsens} and \ref{defn:uwspec}.
Finally, in each run we compute the area under the Receiver Operating Characteristic curves (AUROC).
We compute weighted version of the AUROC by collecting the values of $(\widehat{SN}, \widehat{SP})$ over a grid of decision thresholds ranging from 0 to 1. 


Table \ref{table:comp} summarizes the results over the 500 runs. For sensitivity and specificity at a 50\% decision threshold, the averages of the weighted estimates are closer to the averages of $SN$ and $SP$ than the unweighted estimates.  In particular, an analyst interpreting the unweighted estimates would be overly optimistic on specificity but unduly pessimistic on sensitivity; this is not the case for $\widehat{SN}$ and $\widehat{SP}$. For AUROC, the averages for both the weighted and unweighted estimates are not substantially different from the average population AUROC.  We also see this in the analyses of the NSDUH and NCS-R, as we discuss later.
Overall, the results suggest that using survey-weighted metrics, especially at commonly used decision thresholds, can provide better indications of how a model would perform in the population than unweighted metrics.


\begin{table}[t]
\centering
\begin{tabular}{llll} 
\hline
Metric          & Population     & Unweighted     & Weighted        \\ 
\hline
\multicolumn{4}{l}{Logistic regression}                              \\ 
$\,\,\,\,$Sensitivity & 0.477 ($\pm$ 0.0004)  & 0.426 ($\pm$ 0.001) & 0.478 ($\pm$ 0.001)  \\ 
$\,\,\,\,$Specificity & 0.914 ($\pm$  0.0001) & 0.940 ($\pm$ 0.0002) & 0.914 ($\pm$ 0.0004)  \\ 
$\,\,\,\,$AUROC       & 0.814 ($\pm$ 0.00002)  & 0.825 ($\pm$ 0.0005) & 0.815 ($\pm$ 0.0004)  \\ 
\hline
\multicolumn{4}{l}{Random forest}                                    \\ 
$\,\,\,\,$Sensitivity & 0.460 ($\pm$ 0.0005)  & 0.399 ($\pm$ 0.001) & 0.444 ($\pm$ 0.001)  \\ 
$\,\,\,\,$Specificity & 0.925 ($\pm$ 0.0002)  & 0.944 ($\pm$ 0.0003) & 0.922 ($\pm$ 0.0004)  \\ 
$\,\,\,\,$AUROC       & 0.812 ($\pm$ 0.00006)  & 0.807 ($\pm$ 0.005) & 0.799 ($\pm$ 0.0005)  \\ 
\hline
\multicolumn{4}{l}{Balanced random forest}                           \\ 
$\,\,\,\,$Sensitivity & 0.670 ($\pm$ 0.0004)   & 0.602 ($\pm$ 0.001) & 0.650 ($\pm$ 0.001)  \\ 
$\,\,\,\,$Specificity & 0.800 ($\pm$ 0.0003)   & 0.842 ($\pm$ 0.0005) & 0.791 ($\pm$ 0.0006)  \\ 
$\,\,\,\,$AUROC       & 0.808 ($\pm$ 0.00007)  & 0.804 ($\pm$ 0.0004) & 0.792 ($\pm$ 0.0005)  \\
\hline
\end{tabular}
\caption{\label{table:comp} Results from NSDUH simulation.  Entries are average values over 500 independent runs.  Monte Carlo standard deviations are in parentheses.}
\end{table}

\subsection*{Applications to surveys}
\label{sub:genuine}


We next turn to analyzing two genuine surveys without constructing populations. They comprise $38,813$ individuals from the 2017 edition of the NSDUH and $9,205$ individuals from the NCS-R. While we do not have population truths as benchmarks, we can compare the weighted and unweighted performance metrics. We also estimate the standard errors of the ratio estimators in \eqref{SNhat} and \eqref{SPhat} using the ``survey'' package in R \cite{lumley}.

Unlike our previous simulation,  
these surveys are not stratified samples; rather, they use multi-stage sampling designs.  With the NSDUH data, we use the outcome and predictors described previously.
For the NCS-R data, we use the outcome and predictors from analyses done by Jiang et al. \cite{jiangAddressingMeasurementError2021a}. They  use random forests to predict whether or not someone has attempted suicide from an array of mental illness and substance use variables, including  panic disorder, agoraphobia, specific phobia, social phobia, post-traumatic stress disorder, major depressive disorder, alcohol abuse, alcohol dependence, drug abuse, and drug dependence. Among the $9,205$ individuals in the sample, $8,794$ have not attempted suicide, while $411$ have, making this an outcome variable with significant imbalance in the classes.

For both analyses, we proceed with 50 train-test cycles using balanced random forests. That is, in each of 50 runs, we use simple random samples to split the survey data into training and test sets comprising 80\% of records and 20\% of records, respectively. We use a standard upsampling approach~\cite{japkowicz2002class} to make each $D_t$, which we use to train the random forest using 
1,000 trees and the Gini impurity splitting criteria. We compute weighted and unweighted estimates of the sensitivities, specificities, and AUROCs on each $D_e$. The 50 runs serve a different purpose than those in the previous simulation. 
Specifically, we use the 50 train/test splits to assess the stability of the estimated performance metrics over different splits.
In practice, the analyst may well 
train the predictive model using a single train/test split. 



\begin{table}[t]
\centering
\begin{tabular}{lll}
\hline
                 & Unweighted & Weighted \\ \hline
NSDUH analysis \\
$\,\,\,\,$Sensitivity & 0.71 ($\pm$ 0.003)           &  0.63 ($\pm$ 0.004)        \\ 
$\,\,\,\,$Specificity & 0.76 ($\pm$ 0.001)           &  0.81 ($\pm$ 0.001)        \\ 
$\,\,\,\,$AUROC       & 0.80 ($\pm$ 0.001)           &  0.80 ($\pm$ 0.001)        \\ \hline
NCS-R analysis\\
$\,\,\,\,$Sensitivity & 0.66 ($\pm$ 0.02)           &  0.67 ($\pm$ 0.02)        \\ 
$\,\,\,\,$Specificity & 0.72 ($\pm$ 0.01)           &  0.73 ($\pm$ 0.01)        \\ 
$\,\,\,\,$AUROC       & 0.73 ($\pm$ 0.003)           &  0.74 ($\pm$ 0.004)        \\ \hline
\end{tabular}
\caption{Weighted and unweighted performance metrics for the NSDUH and NCS-R analyses.  Monte Carlo standard deviations are in parentheses.}
\label{tab:svd-NSC-R}
\end{table}

Table \ref{tab:svd-NSC-R} displays results for the NSDUH and NCS-R analyses. The differences in the unweighted and weighted metrics vary by survey. With the NSDUH data, the differences are large, with $\widehat{SN}$ nearly 0.08 lower than the unweighted estimate of sensitivity and $\widehat{SP}$ nearly $0.05$ higher than the unweighted estimate of specificity. In contrast, in the NCS-R analysis, the weighted and unweighted metrics have nearly identical values. For both surveys, the Monte Carlo standard deviations indicate that the variations across the 50 runs are small, suggesting the results are stable.

\begin{figure}
\centering
\begin{subfigure}[b]{0.4\textwidth}
  	\includegraphics[width=\textwidth]{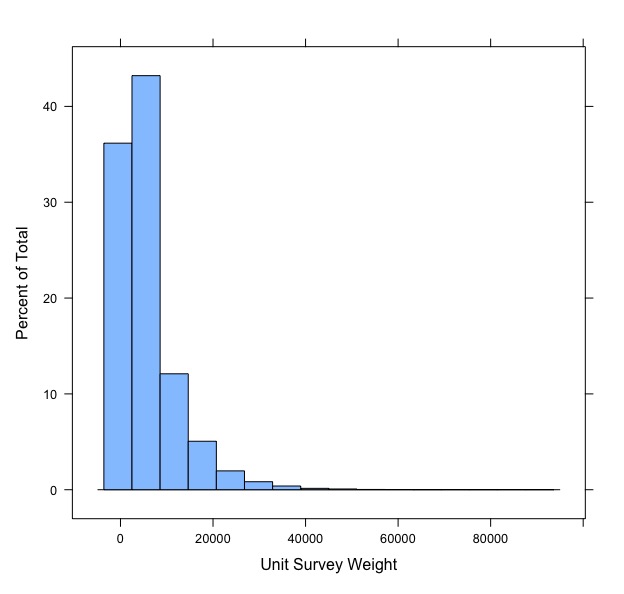}
  	\caption{NSDUH weight distribution}
  	\label{fig:nsduh-dist}	
\end{subfigure}
	 \hfill
\begin{subfigure}[b]{0.4\textwidth}
  \includegraphics[scale=0.3]{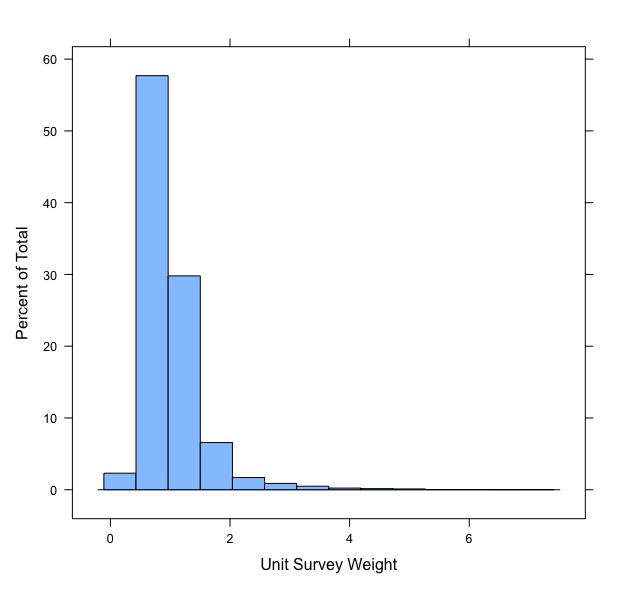}
  \caption{NCS-R weight distribution}
  \label{fig:ncs-dist}
\end{subfigure}
\caption{Weight distributions in the two surveys used in the genuine data analyses.} 
\end{figure}

To understand the contrast in results for the NSDUH and NCS-R analyses, we compare the distributions of the survey weights in the two surveys; these are displayed in Figure \ref{fig:nsduh-dist} and Figure \ref{fig:ncs-dist}. 
The coefficient of variation for the weight distribution in the NSDUH data  is 1.09, whereas for the NCS-R data it is 0.52,  
Evidently, the deviations in individuals' weights from their average are much larger in the NSDUH than in the NCS-R.  
When weights are dispersed, as in the NSDUH, we expect  weighted estimates to differ more noticeably from their corresponding unweighted estimates. When weights are close together, as in the NCS-R, weighted estimates are not much different than multiplying each observation's indicator variable in \eqref{TPest}--\eqref{FNest} by a constant, which cancels from the numerator and denominator of the ratio estimators used to make $\widehat{SN}$ and $\widehat{SP}$.   

For both surveys, the weighted and unweighted averages of AUROC are nearly identical. We conjecture that this arises because the weighted and unweighted values of sensitivity and specificity are not likely to differ at 
threshold values far from the predicted probabilities generated by the classifier.  
For example, suppose a classifier mostly gives predicted probabilities near 0.5. Suppose we consider the threshold for classifying someone as having a disease as a $95\%$ predicted probability or higher.  For most individuals, $TP_i$ in \eqref{tpi} likely equals zero, since few individuals if any will have a large enough predicted probability to be classified by the model as  having the disease. Whether or not one multiplies values of $TP_i$ by $w_i$ does not strongly affect the estimated sensitivity value, which we expect to be low.
We expect the opposite to hold for very low decision thresholds as well. 
Thus, in data where many threshold values are outside the range of most predicted probabilities, we do not expect the use of weights to have noticeable impact on the AUROC.


For any run, we also can compute standard errors for the survey-weighted, estimated population quantities.  To give a sense of the magnitudes of the estimated standard errors, for each analysis we compute typical estimated standard errors as the square root of  the average of the 50 estimated variances obtained from the ``survey'' package.  For the NSDUH analysis, these averages are $0.02$ and $0.007$ for the  sensitivity and sensitivity, respectively.  For the NCS-R analysis, these averages are $0.06$ and $0.01$ for the sensitivity and specificity, respectively. Thus, we expect the evaluation metrics to be estimated reasonably accurately, with the highest uncertainty about the sensitivity in the NCS-R.  Of note, these estimated standard deviations also reflect uncertainty from sampling $D$, which is not present in the Monte Carlo standard deviations of Table \ref{tab:svd-NSC-R}.



\section*{Discussion}
\label{sec:discussion}

 By defining finite population quantities for sensitivity and specificity, analysts can use techniques from survey sampling to estimate how a classifier would perform if applied to the full population.  The survey-weighted estimators have appealing theoretical properties and perform well in the simulation studies presented here.  They can be used in typical data science pipelines that split data into training and test sets. They also can be used with any classifier; for example, 
 an analyst can employ methods to mitigate  class imbalance, and use the survey weights in the model assessment phase to gauge population-level accuracy. More broadly, the analyst can use whatever steps needed to improve in-sample accuracy of the classifier, and so long as survey weights are used when assessing predictive performance in the test set, the accuracy measures reflect how the model predictions would perform if applied to all units in the population.  Given the findings presented here, we recommend that analysts use weighted metrics when reporting results from machine learning studies that make use of survey data.

\newpage

\printbibliography

\end{document}